\documentstyle[epsf,aps,prb,preprint]{revtex}
\begin{document}
\newcommand{\eps}{\varepsilon}  
\newcommand{\vare}{\varepsilon}  
\newcommand{\vareps}{\varepsilon}  
\newcommand{\De}{$\Delta$}
\newcommand{\de}{$\delta$}
\newcommand{\mc}{\multicolumn}
\newcommand{\be}{\begin{eqnarray}}
\newcommand{\ee}{\end{eqnarray}}
\newcommand{\einf}{\varepsilon^\infty}
\newcommand{\ez}{\varepsilon^0}  

\draft \title{Macroscopic polarization and band offsets at nitride 
heterojunctions}
\author{
Fabio Bernardini and Vincenzo Fiorentini}

\address{
Istituto Nazionale di Fisica della Materia - Dipartimento di Scienze
Fisiche, Universit\`a di Cagliari, I-09124 Cagliari, Italy }

\date{Received \ \ \ \ \ \ \ \ \ }
\maketitle

\begin{abstract}
Ab initio electronic  structure studies of prototypical polar
interfaces  of wurtzite III-V nitrides show that  
 large uniform electric fields exist 
 in epitaxial nitride overlayers, due to 
 the discontinuity   across the  interface of the macroscopic
polarization of the constituent materials.
Polarization fields forbid a standard evaluation of band offsets and
formation energies: using new techniques, we find a large
forward-backward asymmetry of the 
offset (0.2 eV for AlN/GaN (0001), 0.85 eV for
GaN/AlN (0001)), and tiny interface formation energies.
\end{abstract}

 % Electr. states in low dim structures

%\begin{multicols}{2}
Due  to their low-symmetry crystal structure, wurtzite
III-V nitrides exhibit  a non-zero macroscopic polarization even in 
equilibrium (spontaneous polarization).\cite{noi.piezo}
Because of the appreciable lattice   mismatch between  nitrides, and
of the fact that nitride heterostructures are usually grown along the
polar (0001) axis, the macroscopic polarization in an epitaxially
grown nitride layer will include a piezoelectric term. Piezoelectric 
constants \cite{noi.piezo} much
larger than in  most other semiconductors imply that small  strains can
produce unusually  large polarizations in III-nitrides. Also,
 spontaneous and piezoelectric  polarizations are  comparable in
magnitude.\cite{noi.piezo} Therefore, a major influence of
polarization on interface and device properties should be
anticipated. 

In this paper we present a detailed first-principle density functional
 theory study (with full account of strain and polarization
 effects) of a prototypical  strained, polar, wurtzite nitride
 interface: GaN/AlN (0001). The central results discussed below are {\it
 (i)}  the change in macroscopic polarization across the
 heterointerfaces generates large uniform electric fields  in the
 layers composing the nanostructure, and {\it (ii)} a large
 forward-backward  band offset asymmetry exists, due to the effects of
 epitaxial strain   on the bulk band structure.
While analogous (though much smaller) fields have 
 been previously predicted in strained superlattices of
 zincblende compounds and in ordered III-V  alloys,\cite{zunger} 
 III-V nitrides stand alone because  of
 their unusually strong polarization,\cite{noi.piezo} 
 both spontaneous {\it and}  piezoelectric.
The presence of  large polarization fields has a host of
interesting consequences on device design which will be discussed in
detail elsewhere.\cite{noi.device} 

Most investigations so far have focused on the interface
 band offset and its possible  asymmetry 
(the offset for AlN on GaN may differ from that of GaN on AlN). It is
 clear   that   measurements and theoretical predictions of
 this basic ingredient of heterostructure design may  be significantly
influenced by macroscopic  polarization and by  strain effects (both
direct on the band bulk structures, and indirectly through
piezoelectric effects). 
 Surprisingly, apart from notable
 exceptions,\cite{noi.mrs,Nardelli}  the recent  experimental
 \cite{Waldrop,Martin} and theoretical \cite{Albanesi,Wei,Walle}
 literature in this field has either disregarded or  analyzed in
 insufficient detail  the effects  of macroscopic bulk polarization on
 interface electronic structure. 
In particular,    theoretical work so far mostly  dealt with
 zincblende\cite{Albanesi,Walle} or 
 artificially lattice-matched wurtzite\cite{Wei} interfaces,
  thus implicitly
 disregarding part of the key ingredients of the problem (polarization,
 or strain, or a combination of both).

Technical details of the
local-density-functional ultrasoft-pseudopotential \cite{uspp}
plane-wave technique and of  the theory of polarization \cite{KS}
employed here are reported in recent  papers.
\cite{noi.piezo,noi.device,noi.mrs,noi.eps} 
Results on bulk lattice parameters,\cite{noi.piezo} dielectric
\cite{noi.eps} and piezoelectric \cite{noi.piezo} constants, 
and spontaneous polarization have also been reported previously.
Technicalities specific
to interface calculations will be reported elsewhere.\cite{noi.device}
Here we only  mention that we  accurately reproduced  previously reported
studies \cite{Wei,Walle} for GaN/AlN (111) interfaces, and that our
results for  GaN/AlN (0001) are in good agreement (where they can be compared)
with similar calculations by a different group.\cite{Nardelli}

Here we study  (GaN)$_n$/(AlN)$_m$(0001) superlattices 
 such that  internal fields do not cause metallization
and at the same time the repeated interfaces are fully decoupled
($n$=$m$=4). Polarization effects on arbitrary  nitride quantum
 structures will be discussed in Ref.\onlinecite{noi.device}. 
We impose to the superlattice the in-plane lattice constant of either
 GaN or AlN in order to simulate the epitaxial relation  of an
 hetero-overlayer on either a GaN or an AlN substrate. The axial
 lattice parameter and internal parameters of the epitaxial material
 are optimized at the imposed subtrate in-plane  lattice parameter. 

We evaluate the valence band offset by splitting it  conventionally
 \cite{Baldereschi} into
 the difference  $\Delta E_v$ of the  bulk  valence band energies
for the two bulks, and the interface potential
line-up $\Delta V$. The latter is generally just a jump in
potential across the interface  from one constant value to another.
Our first result is that this is not the case at polar nitride
 interfaces, and that   the line-up 
cannot be obtained in a conventional fashion.
Indeed, consider Fig. \ref{fig.scf},  which   show
 the macroscopic average \cite{Baldereschi} of the total charge
 density, and the ensuing electrostatic
 potential, of a GaN-matched GaN/AlN (0001) superlattice.
The foremost unusual feature is of course the presence, in the 
 bulk-like regions between  the interfaces,
 of very large ($\sim$ 10$^9$ V/m) uniform electric  fields  generated
 by the different  charge distributions at the 
two interfaces (the  density vanishes
far from the interfaces, which indicates that  the bulk-like regime is 
reached in our simulation).

The main  consequences are: {\it (a)} 
the difference between the  bulk values of the  electrostatic
potential at the two sides  of the interface cannot be defined
unambiguously,\cite{pippa} as it will depend on the choice of the
interface position or of the center of the bulk-like region, which are
of course ill-defined;   {\it (b)} because of energy contributions 
 due to  strain and the electrostatic field,  the formation energy 
 cannot be  extracted as a 
straightforward difference between  total energies and chemical
potentials. In addition {\it (c)}  the origin of the
interface charge asymmetry must be identified; we will show that
an interface charge accumulation takes place because of the
discontinuity of the macroscopic (spontaneous and piezoelectric)
polarization across the interface.

We now show that the determination of the potential
line-up (point 
{\it (a)}) and the identification of the sources of the uniform fields
(i.e., charge asymmetry, point {\it (c)}) can be obtained via  a 
{\it multipole decomposition} of the
 macroscopically-averaged interface charge density. The latter
contains  multipoles of all order, which in 
one-dimensional space  are its moments.
We are interested in the  constant potential drop across the
interface: this is  uniquely determined 
 by the interface {\it dipole}.\cite{nota} We are also interested in
understanding the $\vee$-shaped superlattice potential: these are of course
generated by the 
 interface  {\it monopole}.\cite{cazzo}
All higher multipoles do not generate any potential jumps or
uniform fields, but only minor
 potential bumps at the interface, symmetric
and  antisymmetric  for even and odd multipoles respectively.  
Therefore, in practice, to extract the effects of
 monopoles and dipoles, we simply need to 
 decompose the total macroscopically-averaged charge
density $\bar{\bar{n}}$  into two components comprising,
respectively, all its even  and odd
multipoles.

For the sake of clarity, we name  the odd and even components
respectively  the
 dipole density $\bar{\bar{n}}_{\rm dip}$,
 and the monopole
density $\bar{\bar{n}}_{\rm mono}$. This is admissible since these 
densities produce all the effects of  dipolar and monopolar charges
relevant to our problem,
plus other minor effects related to higher multipoles (irrelevant for
our purposes).

Unfortunately, such a decomposition can be done in
an infinite number of ways.
 Our procedure to obtain $\bar{\bar{n}}_{\rm mono}$ 
is to fold the density with respect to a  mirror plane placed at a
point $z_0$ 
roughly halfway between two adjacent interfaces, and then perform an
antisymmetric combination of  the two charge distributions thus
superimposed, i.e. 
\be 
\bar{\bar{n}}_{\rm mono}
(z-z_0) = \frac{1}{2}\left[\bar{\bar{n}}(z-z_0) 
                     - \bar{\bar{n}}(z_0-z)\right] ,   
\ee
where $z_0$ is the position of the folding plane.
The  dipole density $\bar{\bar{n}}_{\rm dip}$ 
is defined as  the difference between  the
full density and  the monopole term,
or (which amounts to the same)  the
 symmetric combination of the two superimposed
densities
\be
\bar{\bar{n}}_{\rm dip}
        (z-z_0) = \frac{1}{2}\left[\bar{\bar{n}}(z-z_0) 
                     + \bar{\bar{n}}(z_0-z)\right] .
\ee
The key point of this procedure is of course  the choice of the 
position $z_0$ for the folding plane, which implicitly selects 
one specific realization of the decomposition.
 Our criterion for choosing $z_0$
is that the {\it norm} of the monopole component,
\be
S(z_0) = \int |\bar{\bar{n}}_{\rm mono}
(z-z_0)|^{\,2}\,  dz\, ,   
\ee
 should be minimized. This choice
produces {\it (a)}  a dipole distribution which deviates {\it 
minimally} in a least-squares sense from the total density;
{\it (b)} a $\bar{\bar{n}}_{\rm mono}$
  optimally localized at the
interface;
{\it (c)}  a position for
the folding plane
that coincides with the intuitively appealing idea of midpoint between
adjacent interfaces.

The monopole and dipole distributions obtained by the above
decomposition are  shown in Fig.~\ref{fig.mono} for a typical
case. The dipole is 
 related to a jump in potential across the interface, and it
allows the direct determination of the line-up potential, and
therefore of the band offset. In turn, the interface monopole
can be further analyzed to ascertain its physical origin.

Let us first present the valence-band offset of the
GaN/AlN (0001) interface.  The offset is of type I.
As reported  in Table~\ref{tab.vbo},
we obtain $\Delta E_v$=0.20 eV for AlN  lattice matched to GaN, and
 $\Delta E_v$=0.85 eV for GaN 
matched to AlN. We thus  confirm the existence of a  large
forward-backward  asymmetry (0.65 eV) of the offset.
This asymmetry originates mostly from strain-induced band edge shifts in
the bulk band structures (contributing 0.47 eV),
with moderate contributions from the lineup term (0.18 eV).
The large band offset asymmetry is thus mostly to be ascribed
to the different nature of the valence band edge in GaN and AlN.
Indeed, the AlN (GaN) band edge is a singlet (doublet) formed by the 
hybridization along the {\it c}-axis (in the {\it a}-plane) 
of N 2{\it s} orbitals with Al {\it p$_z$} (Ga {\it p$_{xy}$})
states, so that biaxial compression pushes the edges upward in GaN and
downward in AlN.
We note that our findings are semiquantitatively in agreement with
those reported by Nardelli {\it et al.} \cite{Nardelli} for zincblende
(001) interfaces, namely 0.44 eV for AlN on GaN, and 0.73 eV for GaN on AlN.

Let us now turn to the  interface
monopole. The dipole is understood to be \cite{Baroni}
 a  response to the electrostatic perturbation induced by interface
formation (for the present system, in which this effect is
adulterated by monopole contributions, our decomposition
gives the best approximation
to this response). On the other hand,
 the monopole may be expected to be the difference in macroscopic  
polarization between the constituents of the junction: indeed,
according to  Poisson's equation, a polarization discontinuity 
at the interface between two different media produces an interface charge
accumulation. In particular, in a superlattice made of alternating
layers of materials $A$ and $B$ of respective thicknesses $l_A$ and $l_B$
and dielectric constants $\varepsilon_A$ and  $\varepsilon_B$,
the areal charge density at the interface is directly connected
\cite{noi.mrs,noi.eps,VKS} with the {\it transverse} bulk
polarizations  $P^T_A$ and $P^T_B$ of the interfaced materials 
by   
\be
\sigma_{\rm int} =  (P^T_B - P^T_A)\,
( l_A + l_B )/ (l_A \varepsilon_B + l_B \varepsilon_A),
\label{eq:sigma}
\ee
where we have assumed conventionally that $P^T_B$ ($P^T_A$)
is the transverse polarization on the right (left) side of the
interface.\cite{nota2}
This relation  allows an {\it  independent prediction} of what the
polarization-induced 
interface monopole should be, which can be compared with the minimal
monopole {\it calculated for the actual interface}.
Fortunately, the transverse  polarization $P^T$ of the nitrides
can be computed accurately \cite{noi.piezo}  by means of the geometric
quantum phase approach 
\cite{KS} in an arbitrary strain state, for instance
for the epitaxially strained overlayer material. The dielectric
constants (static or electronic) can also be evaluated independently
using a recently developed technique.\cite{noi.eps}

For the unrelaxed structure (clamped ions), the
electronic dielectric constant should be used in Eq. \ref{eq:sigma},
as appropriate to purely electronic screening. 
In  the real system, however, the electric field induces  a lattice
distortion which 
extends over the whole slab, i.e. a long-wavelength  optical phonon
gets frozen-in: it is then appropriate to use in Eq.~\ref{eq:sigma}
the static dielectric constant as calculated in our previous
work.\cite{noi.eps}
In Table~\ref{tab.vbo} we report  the actual interface charge density  
$\sigma_{\rm int}^{\rm (SL)}$ obtained via the multipole decomposition,
and the  value $\sigma_{\rm int}^{\rm (\Delta P)}$ obtained from 
Eq.~\ref{eq:sigma}, for  both the ideal and the relaxed superlattice.
The excellent agreement of the pairs of independently determined
values confirms indeed the identification of the interface charge with
a polarization  charge.

 A final  important issue  is the
evaluation of the interface  formation energy.
For the present system it is impossible to build a superlattice
with equivalent interfaces,  so that a total energy
calculation can only provide  an average interface
formation  energy; this is hardly a severe problem, as the two
interfaces are very similar.\cite{nota}  An additional problem is that
the superlattice total energy contains elastic and
electrostatic energy contributions due to,
respectively,  lattice mismatch and 
polarization  fields. Clearly, these contributions are
extensive, i.e. they depend on the overlayer thickness when referring
formation energies to unit area. In analogy to surface energies, 
we write the total energy per superlattice unit cell as 
\be
E_{\rm tot}^{\rm SL} (n^{\rm X}) =  2 E_f^{\rm int} 
+ \sum_{\rm X} n^{\rm X}(\mu^{\rm X}+\xi^{\rm X}+\eta^{\rm X})
\label{evsn}
\ee
where $\mu^X$ are the total bulk energies per  Ga-N or Al-N pair (in the
appropriately strained geometries), $\xi^X$ are the elastic energies  and $\eta^X$
the electrostatic energies stored in the (possibly) strained bulks under
the polarization field, and $n^{X}$ is the number of atom pairs of type
$X$ (GaN or AlN).

In the present case of a strained low-symmetry system, 
 an exact numerical  equivalence of bulk and interface 
(in particular  between k-point meshes) cannot be achieved,
and the use of  $\mu$, $\xi$ and $\eta$  
evaluated from separate bulk calculations might lead to inaccuracies.
A  solution to this issue, as in the case of surfaces, \cite{fm}
  is to recognize
that $E_{\rm tot}^{\rm SL}$  depends linearly on $n^X$, so that  
 $E^f$ can be
 extracted  as the intercept of the linear  $E_{\rm tot}$ vs. $n_X$
 relation, i.e. from a  series of  total energy
calculations for superlattices of different lenghts
 (whereby equivalent k-point  sets are easily obtained).

In Table~\ref{tab.ef} we list the formation energies for
the ideal and relaxed interfaces obtained by linear extrapolation. The
same Table reports bulk values of the elastic and electrostatic
energy, the former obtained as total energy difference with the
unstrained lattice, and the latter as 
\be
\eta^X = \frac{1}{2} \vareps_X \Omega^{\rm X} E^2 
\ee
with $\Omega^{\rm X}$ is the bulk cell volume, $E$ the modulus of
the electrostatic field, and $\vareps_X$ the static dielectric
constant of material $X$ (the static dielectric constant implicitly
accounts for the field-lattice coupling \cite{noi.eps}). The strain
energy is much larger than the interface energy and  the electrostatic
energy, even for modest thicknesses.
Assuming an order of magnitude for the  dislocation core formation
energy of $\sim 0.5$ eV,\cite{antiphase}  we see that  the formation
of such strain-related defects should start  at typical thicknesses of
$\sim 20$ \AA.  A comparable  electrostatic energy would be stored
 in (perfect) layers of thickness in the order of 500 \AA,  and is
therefore irrelevant to the layer's stability, 
since  metallization or screening effects set in at much smaller
thicknesses.\cite{noi.device} Thus, it can be
safely stated that (0001) nitride interfaces are abrupt, and that
the electrostatic energy should not prevent their stability, as it may
in heterovalent systems such as ZnSe/GaAs. 

In summary, our study of
AlN/GaN(0001) interfaces has revealed the presence of large
uniform electrostatic fields which we demonstrated to originate
 from the macroscopic polarization
of the junction constituents. We have also indicated ways of
extracting band offsets and formation energies, for which
 conventional definitions are useless in the present
situation. We found a sizable
forward-backward  band offset asymmetry,  tiny interface
 formation energies, and large epitaxial strain energies.

Partial support by the European Community under
 Contract BRE2-CT93, and by CINECA Bologna via
Supercomputing  Grants is acknowledged.

\narrowtext
%%%%%%%%%%%%%%%%%%%%%%%%%%%%%%%%%%%%%%%%%%
% T A B L E S 
%%%%%%%%%%%%%%%%%%%%%%%%%%%%%%%%%%%%%%%%%%

\begin{table}
\begin{tabular}{ldddd}
Substrate  $\rightarrow$   &\mc{2}{c}{GaN}      &\mc{2}{c}{AlN}    \\ \hline
\De $E_v$     & 0.20 & (0.29)   & 0.85  &(1.00) \\
$\sigma_{\rm int}^{\rm (SL)}$
              & 0.014& (0.029)  & 0.011 &(0.022) \\
$\sigma_{\rm int}^{\rm (\Delta P)}$ 
              & 0.014& (0.028)  & 0.011 &(0.022) \\
\end{tabular}
\caption{Valence-band offset \De $E_v$ (eV)  and
monopole charge $\sigma_{\rm int}$ (C/m$^2$) at AlN/GaN (0001)  
for different epitaxial  matching conditions,
and fully  relaxed superlattices (in parenthesis: unrelaxed 
case).}  
\label{tab.vbo}
\end{table}
%%%%%%%%%%%%%%%%%%%%%%%%%%%%%%%%%%%%%%%%%%%%%%%%%%%%%%

\begin{table}
\begin{tabular}{lddddd}  
Substrate $\downarrow$ & $E^{\rm int}_f$ & $\eta^{AlN}$&  $\eta^{GaN}$&  $\xi^{AlN}$ &
$\xi^{GaN}$ \\ \hline
GaN    & 3.9  & 5.6 & 9.7  & 179  &  --- \\
AlN    & 0.4   & 10.9 & 6.3  & --- & 155   \\
\end{tabular}
\caption{Formation, electrostatic, and elastic energy
for an AlN/GaN superlattice for different substrate choices (meV/cell 
or unit area).}
\label{tab.ef}
\end{table}

%%%%%%%%%%%%%%%%%%%%%%%%%%%%%%%%%%%%%%%%%%
% F I G U R E S 
%%%%%%%%%%%%%%%%%%%%%%%%%%%%%%%%%%%%%%%%%%
\begin{figure}%[h]
%\unitlength=1cm
%\begin{center}
%\begin{picture}(6,6.3)
%\put(-0.6,-0.0){\epsfysize=6.3cm
%\epsffile{./scf_struc.eps}}
%\end{picture}
%\end{center}
\caption{Total (electronic plus ionic)
 density and ensuing electrostatic potential (in Hartree) 
for an AlN/GaN superlattice matched-lattice to GaN.
The magnitude of the fields in the bulk regions is $\sim 10^9$ V/m.}
\label{fig.scf}
\end{figure}
%%%%%%%%%%%%%%%%%%%%%%%%%%%%%%%%%%%%%%%
\begin{figure}[h]
%\unitlength=1cm
%\begin{center}
%\begin{picture}(6,6.3)
%\put(-0.6,-0.0){\epsfysize=6.3cm
%\epsffile{./fold.eps}}
%\end{picture}
%\end{center}
\caption{Full density (dash-dotted), and monopole (solid) and dipole
(dashed) components for the superlattice of Fig.\protect\ref{fig.scf}.} 
\label{fig.mono}
\end{figure}
%\end{multicols}

\begin{thebibliography}{99}
\bibitem{pippa}
The internal fields will of course affect in the very same way
 the position of core levels, which are generally used
as reference levels in all-electron
calculations (see e.g. Ref. \protect\onlinecite{Wei}).
\bibitem{nota}
Different dipoles at inequivalent interfaces  in the same simulation
cell (as in the 
present case) may generate  an asymmetric-sawtooth--shaped 
potential. This appears not to be present in Fig. \ref{fig.scf},
and  we explicitly tested (using a procedure to be discussed
elsewhere \cite{noi.device})
that  the fields due to interface asymmetry are
 zero to numerical accuracy. This implies  a neglibible
 difference between  dipoles at inequivalent interfaces
and therefore essentially identical offsets.
\bibitem{nota2} 
 Eq. \protect\ref{eq:sigma}  fulfills the periodic boundary
conditions $l_A E_A + l_B E_B = 0$, and it satisfies the 
relationship 
$\vareps_A E_A - \vareps_B E_B = -4\pi (P_A - P_B)$
between transverse polarization and electrostatic field. 
See Refs. \onlinecite{VKS} and \onlinecite{noi.mrs}, and the discussion
 for $l_A = l_B$  in Ref. \onlinecite{noi.eps}. Details of the general
 case will appear elsewhere.\cite{noi.device}
\end{thebibliography}
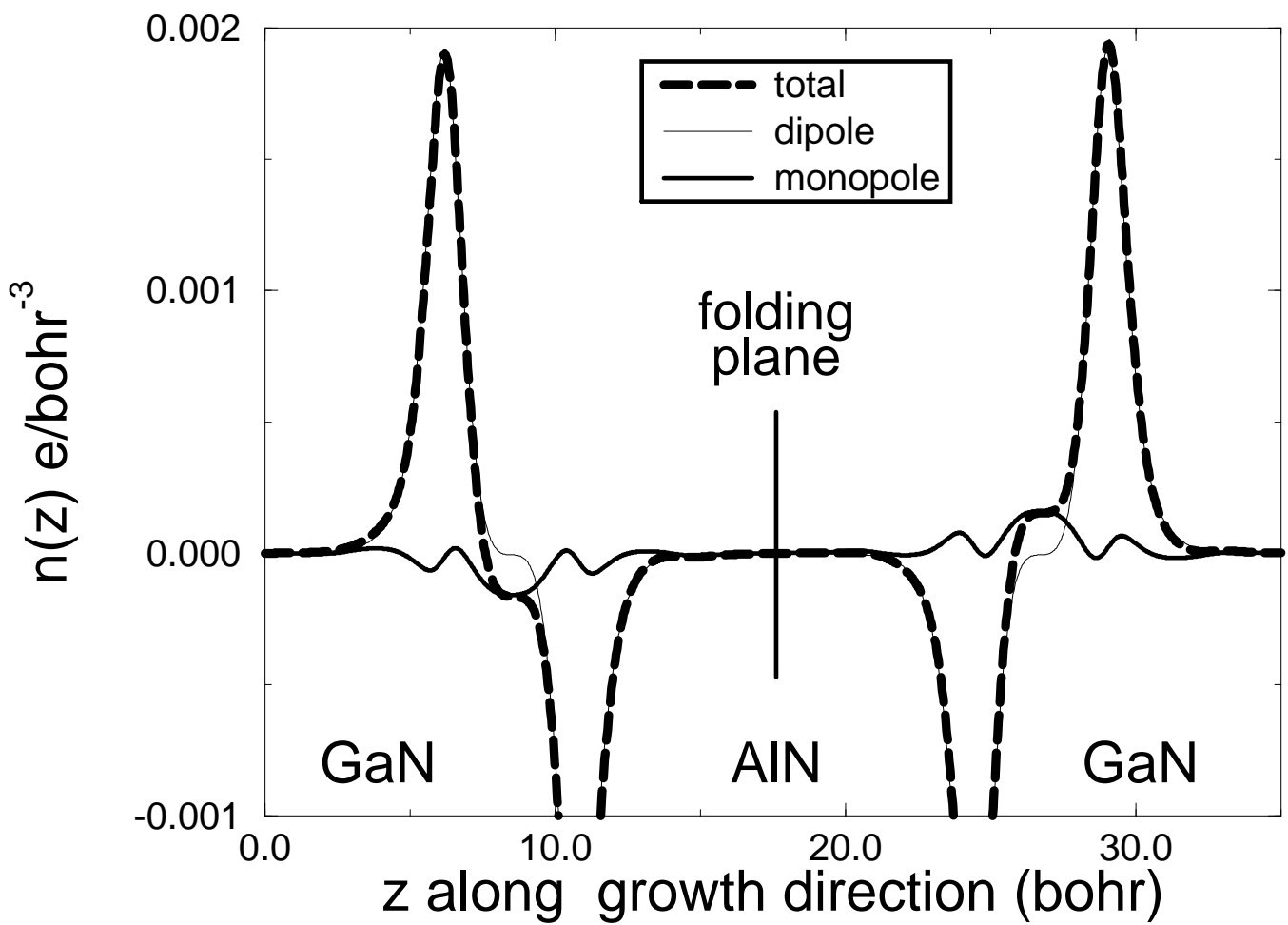
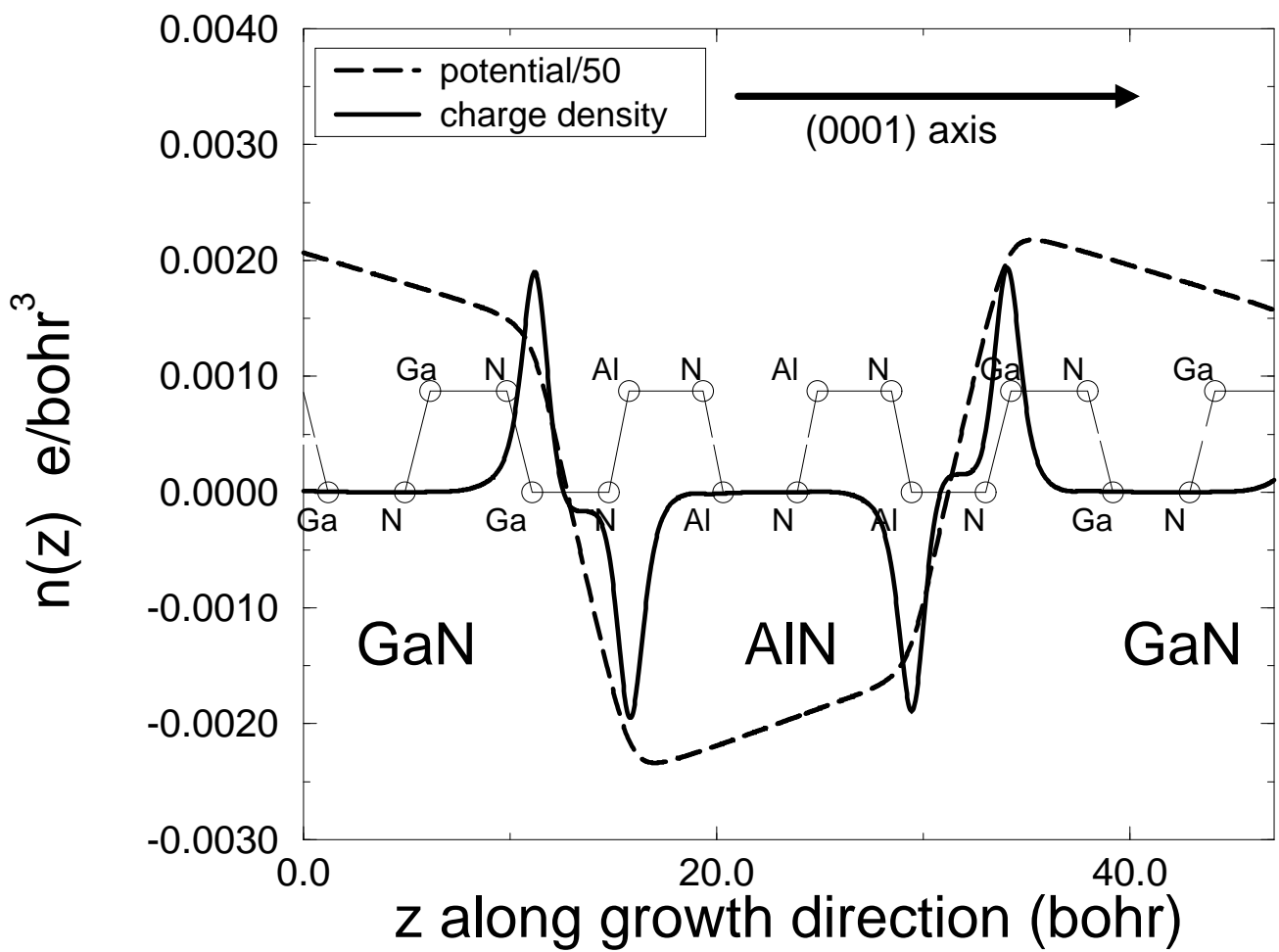
\end{document}